\documentclass[12pt,runningheads,a4paper]{article}
\usepackage{amsmath}
\usepackage{graphicx}
\usepackage{amssymb}
\usepackage[arrow,matrix]{xy}
\usepackage{setspace}
\date{}
\begin{document}
%%%%%%%%%%%%%%%%%%%%%%%%%%%%%%%%%%%%%%%%%%%%%%%%%%%%%%%%%%%%%%%%%%%%%%%%%%%%%%%%%%%%%%%%%%%%%%%%%%%%%%%%%%%%%%%%%%%%%%%%%%%%%%%%%%%%%%%
\title{\textbf{{Effects of thermal radiation and mass diffusion on MHD flow over a vertical plate applying time dependent shear to the fluid }}}
\maketitle
\author\begin{center}N. Shahid${}^{1,}$\footnote{E-mail address: $nshahid@princeton.edu$
}
\vskip0.5cm
\emph{{\scriptsize${}^{1}${Department of Mechanical and Aerospace Engineering, Princeton University, Princeton 08544, New Jersey, USA}}}\end{center}\vskip1cm
%\begin{spacing}{1}
\textbf{Abstract}\\\\{\small{
\indent\indent The present paper studies the effects of thermal radiation and mass diffusion on MHD flow over a vertical plate that applies time dependent shear to the fluid. This study is meant to provide framework for improved thermal system where induced automated shear on chemical (fluid) will increase velocity of fluid and hence enhances the smooth flow. This study will also throw light on an important aspect of controlling temperature of thermal system in the context of emitting thermal radiation. Exact expressions for velocity field, temperature and mass concentration corresponding to the radiative flow of viscous fluid have been calculated. These expressions are obtained by using Laplace transform of corresponding fractional differential equations. The expressions of temperature and mass concentration of fluid have been presented in series form. However, velocity field is presented in the form of integral solutions. All exact expressions satisfy initial and boundary conditions. Some significant limiting cases of fluid parameters and of fractional parameters have been discussed. Two special cases of shear stress; shear stress in the form of Heaviside function and oscillating shear stress have also been taken into account to compare the behavior of fluid motion graphically. An analysis has also been prepared to compare analytical and numerical solutions for concentration of fluid using numerical algorithm. Validity of analytical solutions up to a certain order of accuracy has been established.

%\end{spacing}

\section{\textbf{Introduction}}
\indent \indent In recent times, a great deal of work has been done on studying radiative heat and mass transfer of free convective flows[1-7]. The importance of these flows arise from their applications in industrial and chemical processes, filtration processes as well as in biological and physical processes [8-11]. Many studies reflecting on combined heat  and mass transfer have appeared recently with various physical scenarios. The phenomenon of mass transfer in fluid at rest owes its appearance to concentration gradients in the fluid that cause molecular diffusion. Due to similar nature of conservative heat and mass transfer processes for low mass concentration and low mass transfer, many authors are tended to deal with heat and mass convective flows simultaneously[12-17]. Moreover, as magnetic hydrodynamic (MHD) fluid is used in many engineering and industrial applications, like cooling of metal in nuclear reactors and magnetic control of iron flow in steel industry, etc. [18-21], the study of properties of radiative heat and mass transfer in MHD fluids has become a central topic of present time work.\\
\indent\indent Many authors have studied convective flows over or past vertical plates to contribute towards development of heat sinks, heat exchanger plate used in isothermal chemical reactor and heating blocks. The problem of convective flow past a vertical  oscillating plate was first considered by Soundalgekar [22] and same problem was extended to study mass transfer in fluid by Soundalgekar and Akolkar [23]. Soundalgekar [24] continued his work on this line by studying effects of mass transfer on the flow past an infinite vertical oscillating plate with constant heat flux. Thermal radiation effects on laminar free convection boundary layer of an absorbing  gass and on the combined free and forced convection of electrically conducting fluid in the presence of transverse magnetic field have been studied by  England and Emery [25] and Gupta and Gupta [26]. Many interesting physical aspects of radiative flows  and effects of mass transfer on fluid motion with respect to varying temperatures have been explored by Hosain and Takhar [27], Mazumdar and Deka [28], and Gebhart and Pera [29]. The combined effects of radiation and chemical reaction on free convective flow in porous medium were studied by Deka and Neog [30]. The important thing to bear in mind here is that in all above works, the initial and boundary condition on velocity, shear stress and mixed conditions were employed for various  physical situations but the behavior of fluid motion with condition that the force or shear is applied  on boundary has yet to be explored. The above mentioned literature reflected on the impact of temperature variation of plate and induced magnetic field on velocity of fluid. The added force of translating and oscillating plates also influences the energy transfer processes, was also described in great details. However, to the best of our knowledge, the influence of induced time dependent shear on heat energy and mass transfer applied by vertical plate has not been discussed yet. This aspect of convective flow is explored here. The concerned theoretical findings are expected to enhance smooth transport of heat energy and mass, and accelerate conduction processes while dealing with anomaly of raised system temperature. Waters and King [31], Bandelli et al [32] , as well as some authors  [33-35] have pursued the study of simple fluid motion when shear stress on boundary is specified and have described how this particular boundary condition influences the flow.\\
\indent\indent Furthermore, the complex dynamics of various viscous fluids can be aptly described  by fractional calculus  mainly because fractional constitutive relationship model assesses information about molecular substance more efficiently than customary constitutive relationship models.
Fractional calculus is considered to be an important tool that gives way to achieving generalization of many classical physical results. Its applications include fractional Hamiltonian dynamics [36,37], fractal media [38] and fractional diffusion equation [39], etc. In order to study the properties of viscous fluids, Germant [40] firstly proposed the use of fractional derivatives. Then, Slonimsky [41] described the relaxation process by introducing fractional derivatives into Kelvin-Voigt model. This theory was, then, extended by Bagley and Torvik [42,43] and Koeller [44]. They established the fact that constitutive relations with fractional derivatives predicted the theory of hereditary solid mechanics and the theory of viscoelasticity of coiling polymers. Consistence of fractional derivatives model with basic theories make them more reliable. Many recent contributions [45-54] have been made  using fractional calculus approach to study motion of fluids  for different physical settings.\\
\indent\indent The present work is motivated to improve a thermal system, say a heating block, where induction of automated shear on chemical (fluid) above plate will enhance smooth flow while keeping the temperature in control. In particular, this paper studies the dynamics of fluid model in which transfer of heat energy and mass is influenced by shear applied on fluid by vertical plate. Though, the shearing force does not affect the temperature of fluid but it inreases the flow velocity. Transfer of heat energy is heightened with increase in temperature which in turn increases the thermal radiation. Temperature check of fluid gets important  as increase in temperature beyond certain measure could lead to final product flaw or further complications of chemical reactions. The present study will throw light on how thermal radiation energy could influence the temperature of fluid, how considered geometrical configuration of model impacts the temperature and how conduction procedures could be improved based on relation of temperature of fluid with Prandtl number. Mainly, the purpose of this paper is to determine the factors that would ensure increase in velocity of fluid under the application of time-dependent shear. The obstruction caused by magnetic and viscous forces to prevent smooth flow of fluid will also be discussed and illustrated through mathematical and graphical approach.\\
\indent\indent Here, we obtain exact solutions for unsteady MHD fluid over an infinite plate that applies time dependent shear  $f(t)$ to the fluid. The free convection of flow  is studied using fractional derivatives to describe more naturally the complex dynamics of radiative heat and mass transfer in the flow. Viscous dissipation is assumed to be negligible and species concentration is taken to be very low. The effects of thermal radiation, mass concentration and temperature parameters on free convective flow are studied in the presence of magnetic field. All initial and boundary conditions are satisfied by obtained general solutions with fractional parameters and limiting values of these fractional parameters lead to exact solutions of ordinary differential equations for present model. Some recently obtained exact solutions for convective flow problems with boundary condition on velocity have also been retrieved  by considering limiting values of the fractional parameters, validating our solutions in this paper. Also, many interesting physical aspects of radiative flow with mass transfer have been depicted and verified by graphs. In particular, we have compared velocity profiles for two special cases of shear stress i.e. one case corresponds to fluid motion when plate applies constant shear, $f(t)=1$ to the fluid and in the second case velocity profiles have been drawn for oscillating shear stress, $f(t)=sin(\omega t)$ applied by the plate to fluid.
\section{Mathematical formulation of the problem}
\indent Let us consider the unsteady flow of an incompressible viscous electrically conducting MhD fluid over an infinite vertical plate. The x-axis of the Cartesian coordinate system is taken along the plate in vertical direction and the y-axis is normal to the plate. Initially, the fluid and the plate are at the temperature $T_\infty$ and species concentration $C_{\infty}$. At time $t=0^+$, the plate applies a time-dependent shear stress $h(t)$ to the fluid in the direction of x-axis. Also, at the same time the plate temperature is raised to $T_W$ and species concentration to $C_W$ linearly with time. A uniform magnetic field $B_0$ is applied in the normal direction of plate. It is assumed that magnetic Reynold's number is very small and the induced magnetic field is negligible in comparison to transverse magnetic field. The viscous dissipation and Soret $\&$ Duoffer effects due to lower level of concentration are assumed to be negligible.\\
Above assumptions and Boussinesq's approximation lead to the following set of governing equations of unsteady flow
\begin{equation}
\frac{\partial u(y,t)}{\partial t}=\nu\frac{\partial^2{u(y,t)}}{\partial y^2}+g\beta(T(y,t)-T_\infty)+g\beta^*(C(y,t)-C_\infty)-\frac{\sigma B_0^2}{\rho}u(y,t);\,\,\,\,\,y,t>0
\end{equation}
\begin{equation}
\rho C_P\frac{\partial T(y,t)}{\partial t}=\kappa\frac{\partial^2{T(y,t)}}{\partial y^2}-\frac{\partial q_r(y,t)}{\partial y};\,\,\,\,\,y,t>0
\end{equation}
\begin{equation}
\frac{\partial C(y,t)}{\partial t}=D\frac{\partial^2{C(y,t)}}{\partial y^2};\,\,\,\,\,y,t>0
\end{equation}
and initial and boundary conditions with the assumption of no slip between fluid and plate are
\begin{eqnarray}
u(y,t)=0,\,\,T(y,t)=T_\infty,\,\,C(y,t)=C_\infty,\,\,\,\,\,y\geq 0,\,t=0
\end{eqnarray}
\begin{eqnarray}
{\frac{\partial u(y,t)}{\partial y}}\bigg{|}_{y=0}=\frac{h(t)}{\mu},\,\,T(0,t)=T_W,\,\,C(0,t)=
C_\infty+(C_W-C_\infty)\frac{U_0^2 t}{\nu},\,\,t>0
\end{eqnarray}
\begin{eqnarray}
u(y,t)\rightarrow 0,\,\,T(y,t)\rightarrow T_\infty,\,\,C(y,t)\rightarrow C_\infty \,\,as\,\,y\rightarrow \infty
\end{eqnarray}
where $u(y,t)$,$T(y,t)$, $C(y,t)$,$\nu$, $g$, $\beta$, $\beta^*$, $\kappa$, $q_r$, $C_P$, $\rho$ and $D$ are velocity of the fluid, its temperature, species concentration in the fluid, kinematic viscosity, gravitational acceleration, coefficient of thermal expansion, coefficient of expansion with concentration, thermal conductivity of the fluid,  radiative heat flux, specific heat at constant pressure, density of fluid and mass diffusion coefficient, respectively.\\
Also in equation (5), $\mu=\rho\nu$ is the coefficient of viscosity and the function $h(t)$ satisfies the condition $h(0)=0$.\\
Following Cogly-Vincentine-Gilles equilibrium model based on assumption of optically thin medium with relative low density, we have
\begin{eqnarray}
\frac{\partial q_r(y,t)}{\partial y}=4(T(y,t)-T_\infty)\int_0^\infty K_W \bigg(\frac{\partial {e_b}}{\partial {T}}\bigg)_W d\lambda
=4I^*(T(y,t)-T_\infty)
\end{eqnarray}
where $K_W$ and $e_b$ are absorption coefficient and plank function.\\
Introducing Eq. (7) in Eq. (2), we have
\begin{equation}
\rho C_P\frac{\partial T(y,t)}{\partial t}=\kappa\frac{\partial^2{T(y,t)}}{\partial y^2}-4I^*(T(y,t)-T_\infty);\,\,\,\,\,y,t>0
\end{equation}
To obtain solutions of Eqs. (1), (3) and (8) along with initial and boundary conditions (4), (5) and (6), we first convert these equations in dimensionless form.\\
The following dimensionless quantities have been introduced
\begin{eqnarray}
u^{*}=\frac{u}{U_0},\,\,\,y^{*}=\frac{yU_0}{\nu},\,\,\,t^{*}=\frac{tU_0^2}{\nu},\,\,\, T^*=\frac{T-T_\infty}{T_W-T_\infty}\\ \nonumber
C^*=\frac{C-C_\infty}{C_W-C_\infty},\,\,\,P_r=\frac{\mu C_P}{\kappa},\,\,\,S_c=\frac{\nu}{D},\,\,\,\,G_r=\frac{\rho\beta\nu(T-T_\infty)}{U_0^3}
\\ \nonumber
G_m=\frac{g\beta^*\nu(C-C_\infty)}{U_0^3},\,\,\,\,M=\frac{\sigma B_0^2\nu}{\rho U_0^2},\,\,\,\,F=\frac{4I^*\nu^2}{\kappa U_0^2}
\end{eqnarray}
where $U_0$, $P_r$, $S_c$, $G_r$, $G_m$, $M$ and $F$ are a constant, Prandtl number, Schmidth number, thermal Grashof number, mass Grashof number, Hartmann number and dimensionless thermal radiation parameter, respectively.\\
Using dimension less quantities (9) in governing equations (1), (3) and (8) and dropping $"*"$ notation, we obtain
\begin{equation}
\frac{\partial u(y,t)}{\partial t}=\frac{\partial^2{u(y,t)}}{\partial y^2}+G_rT(y,t)+G_mC(y,t)-Mu(y,t)\,\,\,\,\,;y,t>0
\end{equation}
\begin{equation}
\frac{\partial T(y,t)}{\partial t}=\frac{1}{P_r}\frac{\partial^2{T(y,t)}}{\partial y^2}-\frac{F}{P_r}T(y,t)\,\,\,\,\,;y,t>0
\end{equation}
\begin{equation}
\frac{\partial C(y,t)}{\partial t}=\frac{1}{S_c}\frac{\partial^2{C(y,t)}}{\partial y^2}\,\,\,\,\,;y,t>0
\end{equation}
The corresponding initial and boundary conditions are
\begin{eqnarray}
u(y,0)=T(y,0)=C(y,0)=0,\,\,\,\,;y\geq 0
\end{eqnarray}
\begin{eqnarray}
\frac{\partial u(y,t)}{\partial y}\bigg{|}_{y=0}=\frac{1}{\rho U_0^2}h\bigg(\frac{t^*\nu}{U_0^2}\bigg)=f(t),\,\,T(0,t)=1,\,\,C(0,t)=t\,\,\,\,;t>0&\\\nonumber
u(y,t),T(y,t), C(y,t)\rightarrow 0\,\,\, as\,\,\,y\rightarrow \infty
\end{eqnarray}
To obtain analytical formulas for velocity, temperature and concentration, we use fractional derivative approach. In particular, we consider Caputo fractional derivative operator. Equations (10), (11) and (12) with Caputo differential operator take the form
\begin{equation}
D_t^\alpha u(y,t)=\frac{\partial^2{u(y,t)}}{\partial y^2}+G_rT(y,t)+G_mC(y,t)-Mu(y,t)\,\,\,\,\,;y,t>0
\end{equation}
\begin{equation}
D_t^\beta T(y,t)=\frac{1}{P_r}\frac{\partial^2{T(y,t)}}{\partial y^2}-\frac{F}{P_r}T(y,t)\,\,\,\,\,;y,t>0
\end{equation}
\begin{equation}
D_t^\gamma C(y,t)=\frac{1}{S_c}\frac{\partial^2{C(y,t)}}{\partial y^2}\,\,\,\,\,,y,t>0
\end{equation}
where Caputo differential operator $D_t^\alpha$ is defined as [55,56]
\begin{eqnarray*}
D_t^\alpha f(t)=\frac{1}{\Gamma(1-\alpha)}\int_o^t{\frac{f'(\tau)}{(t-\tau)^\alpha}d\tau};\,\,\,0<\alpha<1
\end{eqnarray*}
where $\Gamma(.)$ is the Gamma function.\\
\section{Analytical solutions}
Analytical solutions will be obtained by means of Laplace transform and inverse Laplace transform.\\
Applying Laplace transform to Eq. (17) and using Laplace transform of corresponding initial and boundary condition (13) and (14), we obtain
\begin{equation}
\bar{C}(y,q)=\frac{1}{q^2}e^{-\sqrt{S_cq^\gamma}y}
\end{equation}
where $\bar{C}(y,q)$ is the Laplace transform of $C(y,t)$.\\
In order to obtain $C(y,t)$, we write Eq. (18) in the form
\begin{equation}
\bar{C}(y,q)=\frac{1}{q^2}+\frac{1}{q^2}\sum_{n=1}^\infty\frac{(-\sqrt{S_c}y)^n}{n!}q^{\frac{\gamma n}{2}}
\end{equation}
Applying Laplace inverse transform to Eq. (19), we obtain
\begin{equation}
C(y,t)=t+\sum_{n=1}^\infty\frac{(-\sqrt{S_c}y)^n}{n!}\frac{t^{\frac{-\gamma n}{2}+1}}{\Gamma(2-\frac{\gamma n}{2})}
\end{equation}
satisfying initial and boundary conditions for mass concentration of the fluid.\\
Now, applying Laplace transform to Eq. (16) and using Laplace transform of corresponding initial and boundary conditions (13) and (14), we obtain
\begin{equation}
\bar{T}(y,q)=\frac{1}{q}e^{{-\sqrt{P_rq^\beta+F}y}}
\end{equation}
To find $T(y,t)=L^{-1}\{\bar{T}(y,q)\}$, we firstly write Eq. (21) in the following form
\begin{equation}
\bar{T}(y,q)=\frac{1}{q}+\frac{1}{q}\sum_{n=1}^\infty\frac{(-\sqrt{F}y)^n}{n!}\sum_{j=0}^\infty
\frac{\Gamma(\frac{n}{2}+1)(\frac{P_r}{F})^jq^{\beta j}}{j!\Gamma(\frac{n}{2}-j+1)}
\end{equation}
Taking Laplace inverse transform of Eq. (22), we obtain
\begin{equation}
T(y,t)=1+\sum_{n=1}^\infty\frac{(-\sqrt{F}y)^n}{n!}
\sum_{j=0}^\infty\frac{\Gamma(\frac{n}{2}+1)(\frac{P_r}{Ft^\beta})^j}{j!\Gamma(\frac{n}{2}-j+1)\Gamma(1-\beta j)}
\end{equation}
satisfying initial and boundary conditions of temperature.\\
We can also write the above expression in terms of Fox-H function,
\begin{eqnarray}
T(y,t)=1+\sum_{n=1}^\infty\frac{(-\sqrt{F}y)^n}{n!}
H^{1,1}_{1,3}\left[\frac{-P_r}{Ft^\beta}\bigg{|}\begin{array}{l}(-\frac{n}{2},0)\\
(0,1),(-\frac{n}{2},-1),(0,-\beta)
\end{array}\right]
\end{eqnarray}
where Fox- H function is defined as[57]
\begin{eqnarray*}
\mathop\sum_{n=0}^{\infty}\frac{(-z)^n\prod_{j=1}^{p}\Gamma(a_j+A_jn)}{n!\prod_{j=1}^{q}\Gamma(b_j+B_jn)}=
H^{1,p}_{p,q+1}\left[z\bigg{|}\begin{array}{l}
                                                              (1-a_1,A_1),...,(1-a_p,A_p) \\
                                                              (0,1),(1-b_1,B_1),...,(1-b_q,B_q)
                                                            \end{array}
\right].
\end{eqnarray*}
To find the exact expression for velocity field $u(y,t)$, we apply discrete Laplace transform to Eq. (15) and obtain
\begin{equation}
\frac{\partial^2{\bar{u}(y,q)}}{\partial y^2}-(q^\alpha+M)\bar{u}(y,q)=-G_r\bar{T}(y,q)-G_m\bar{C}(y,q)
\end{equation}
where $\bar{u}(y,q)$ is the Laplace transform of $u(y,t)$.
Also, $\bar{u}(y,q)$ has to satisfy the condition
\begin{equation}
\frac{\partial \bar{u}(y,q)}{\partial y}\bigg{|}_{y=0}=F(q)
\end{equation}
where $F(q)$ is Laplace transform of $f(t)$.\\
Solving Eq. (25) with the help of Eqs. (18), (21) and (26), we obtain
\begin{eqnarray}
\bar{u}(y,q)=&\frac{-F(q)e^{-\sqrt{q^\alpha+M}y}}{\sqrt{q^\alpha+M}}+\frac{G_r\sqrt{P_r q^\beta+F}e^{-\sqrt{q^\alpha+M}y}}{q\sqrt{q^\alpha+M}[P_r q^\beta-q^\alpha+(F-M)]}
+\frac{G_m\sqrt{S_c q^\gamma}e^{-\sqrt{q^\alpha+M}y}}{q^2[S_cq^\gamma-q^\alpha-M]\sqrt{q^\alpha+M}}\\ \nonumber
&-\frac{G_re^{-\sqrt{P_rq^\beta+F}y}}{q[P_r q^\beta-q^\alpha+(F-M)]}-\frac{G_me^{-\sqrt{q^\gamma S_c}y}}{q^2[S_cq^\gamma-q^\alpha-M]}
\end{eqnarray}
To find $u(y,t)=L^{-1}\{\bar{u}(y,q)\}$, we firstly write Eq. (27) in a more suitable form as follows
\begin{eqnarray}
&\bar{u}(y,q)=\frac{-F(q)e^{-\sqrt{q^\alpha+M}y}}{\sqrt{q^\alpha+M}}+
G_rP_r\bigg(\frac{e^{-\sqrt{q^\alpha+M}y}}{\sqrt{q^\alpha+M}}\frac{q^{\beta-1}}{\sqrt{P_rq^\beta+F}}\bigg)\bigg(\frac{1}{P_r q^\beta-q^\alpha+(F-M)}\bigg)\\ \nonumber
&+G_r F\bigg(\frac{e^{-\sqrt{q^\alpha+M}y}}{\sqrt{q^\alpha+M}}\frac{q^{-1}}{\sqrt{P_rq^\beta+F}}\bigg)\bigg(\frac{1}{P_r q^\beta-q^\alpha+(F-M)}\bigg)
+\sqrt{S_c}G_m\bigg(\frac{e^{-\sqrt{q^\alpha+M}y}}{\sqrt{q^\alpha+M}}q^{\frac{\gamma}{2}-2}\bigg)\bigg(\frac{1}{S_cq^\gamma-q^\alpha-M}\bigg)\\ \nonumber
&-\frac{G_re^{-\sqrt{P_rq^\beta+F}y}}{q[P_r q^\beta-q^\alpha+(F-M)]}-\frac{G_me^{-\sqrt{q^\gamma S_c}y}}{q^2[S_cq^\gamma-q^\alpha-M]}
\end{eqnarray}
Applying Laplace inverse transform to Eq. (28) and using Appendix $A_1$, $A_2$, $A_3$, $A_4$ and $A_5$, we obtain analytic expression of velocity field
\begin{eqnarray}
u(y,t)=u_t(y,t)+u_{TC}(y,t)
\end{eqnarray}
where
\begin{eqnarray}
u_t(y,t)=-\int^t_0\int^\infty_0\frac{e^{-\frac{y^2}{4u}-Mu}}{\sqrt{\pi u}}h(u,s)f(t-s)duds
\end{eqnarray}
represents velocity field corresponding to time dependent shear stress and
\begin{eqnarray}
&u_{TC}(y,t)=\frac{1}{P_r}\sum_{p=0}^\infty\frac{1}{(P_r)^p}\int^t_{0}\int^{s^{'}}_{0}\int^\infty_{0}\frac{e^{-\frac{y^2}{4u}-Mu}}{\sqrt{\pi u}}h(u,s)
\bigg\{G_r\sqrt{P_r}G_{\beta,\beta-1,\frac{1}{2}}(\frac{-F}{P_r}, s^{'}-s)
\\ \nonumber
&+\frac{G_rF}{\sqrt{P_r}}G_{\beta,-1,\frac{1}{2}}\bigg(\frac{-F}{P_r}, s^{'}-s\bigg)\bigg\}
G_{\alpha,p\alpha,p+1}(\frac{-F}{P_r}, t-s^{'})dudsds^{'}
\\ \nonumber
&+\frac{G_m}{\sqrt{S_c}\Gamma{(2-\frac{\gamma}{2})}}
\sum_{m=0}^\infty\frac{1}{(S_c)^m}\int^t_{0}\int^{s^{'}}_{0}\int^\infty_{0}\frac{e^{-\frac{y^2}{4u}-Mu}}{\sqrt{\pi u}}h(u,s)
(s^{'}-s)^{1-\frac{\gamma}{2}}G_{\gamma,\alpha m,m+1}(\frac{M}{S_c}, t-s^{'})dudsds^{'}
\\ \nonumber
&-\frac{G_r}{P_r}\sum_{n=0}^\infty\frac{(-\sqrt{P_r}y)^n}{n!}
\sum_{m=0}^\infty\bigg(\frac{F}{P_r}\bigg)^m\frac{\Gamma(\frac{n}{2}+1)}{m!\Gamma(\frac{n}{2}-m+1)\Gamma(\beta m-\frac{\beta n}{2})}\sum_{p=0}^\infty\frac{1}{(P_r)^p}\int^t_0(t-s)^{\beta m-\frac{\beta n}{2}-1}
\\ \nonumber
 &\times G_{\beta,p\alpha-1 ,p+1}(\frac{M-F}{P_r}, s)ds-\frac{G_m}{S_c}
\sum_{n=0}^\infty\frac{(-\sqrt{S_c}y)^n}{n!\Gamma(2-\frac{\gamma n}{2})}\sum_{m=0}^\infty\frac{1}{(Sc)^m}
\int_0^t(t-s)^{1-\frac{\gamma}{2}}G_{\gamma,\alpha m,m+1}(\frac{M}{S_c}, s)ds
\end{eqnarray}
corresponds to thermal radiation and mass concentration of fluid.\\
In Eqs. (30) and (31), $h(u,t)$ is defined as
\begin{eqnarray*}
h(u,t)=L^{-1}\{e^{-uq^\alpha}\}=\frac{1}{\alpha\Gamma(\alpha)}\sum_{n=0}^\infty\frac{(-u)^n}{(n+1)!\Gamma{(\alpha(n+1))}}\int^t_{0}\int^\infty_{0}
(t-s)^{\alpha-1}J_0(2\sqrt{xs})x^{\alpha(n+1)}dxds
\end{eqnarray*}
where $J_0(.)$ is the Bessel function.
\section{Limiting cases}
For $\alpha,\,\,\,\beta,\,\,\,\gamma\rightarrow 1$ in Eqs. (18), (21) and (27), we can obtain $(y,t)$ solutions of governing equations in ordinary differential operator. Some significant limiting cases have been discussed below.
\subsection{Solution in the absence of magnetic field}
The absence of magnetic field i.e. $M=0$ and assumption of $\alpha,\,\,\,\beta,\,\,\,\gamma\rightarrow 1$ lead to the following expression of velocity field
\begin{eqnarray}
&u(y,t)=-\int^t_0\frac{e^{-\frac{y^2}{4s}}}{\sqrt{\pi s}}f(t-s)ds+\frac{G_r\sqrt{P_r}}{\pi(P_r-1)}\int^t_{0}\int^{s^{'}}_{0}\frac{e^{-\frac{y^2}{4s}-\frac{F}{P_r}(s^{'}-s)-\frac{-F}{P_r-1}(t-s^{'})}}
{\sqrt{s(s^{'}-s)}}dsds^{'}
\\ \nonumber
&+\frac{G_r F}{\pi(P_r-1)}\int^t_{0}\int^{s^{'}}_{0}\int^{t-s^{'}}_{0}\frac{e^{-\frac{y^2}{4s}-\frac{F}{P_r}(s^{'}-s)-\frac{F}{P_r-1}u}}
{\sqrt{s(s^{'}-s)}}dsduds^{'}+\frac{G_m\sqrt{S_c}}{S_c-1}\int^t_{0}erfc(\frac{y}{2\sqrt{s}})(t-s)ds
\\ \nonumber
&-\frac{G_m}{S_c-1}\int^t_{0}erfc(\frac{\sqrt{S_c}y}{2\sqrt{s}})(t-s)ds-\frac{G_r}{P_r-1}\sum_{n=0}^\infty\frac{(-\sqrt{P_r}y)^n}{n!}
\sum_{m=0}^\infty\frac{(\frac{F}{P_r})^m}{m!}\frac{\Gamma(\frac{n}{2}+1)}{\Gamma(\frac{n}{2}-m+1)\Gamma(m-\frac{n}{2})}
\\ \nonumber
&\times\int^t_{0}\int^{s^{'}}_{0}e^{-\frac{F}{P_r-1}s}(t-s^{'})^{m-\frac{n}{2}-1}dsds^{'}
\end{eqnarray}
\subsection{Solution in the case of constant radiative heat flux and $\beta\rightarrow 1$}
Assuming radiative heat flux to be constant along y-direction of plate, F=0 (or $q_r$=constant)and $\beta\rightarrow 1$, we obtain from Eq. (21) the following expression
\begin{equation}
\bar{T}(y,q)=\frac{1}{q}e^{{-\sqrt{P_rq}y}}
\end{equation}
Applying Laplace inverse transform to Eq. (33), we obtain an expression for temperature of the fluid in the absence of thermal radiation i.e.
\begin{equation}
T(y,t)=erfc(\frac{\sqrt{P_r}y}{2\sqrt{t}})
\end{equation}
satisfying also the corresponding boundary condition (14) for temperature where $erfc(.)$ represents complementary error function.
\subsection{Solution in the absence of magnetic field and constant radiative heat flux}
The absence of magnetic field, constant radiative heat flux along y-direction of plate, F=0 (or $q_r$=constant) and assumption of $\alpha,\,\,\,\beta,\,\,\,\gamma\rightarrow 1$ in Eq. (27) lead to the following expression of velocity field
\begin{eqnarray}
&u(y,t)=-\int^t_0\frac{e^{-\frac{y^2}{4s}}}{\sqrt{\pi s}}f(t-s)ds+
\frac{G_r\sqrt{P_r}}{P_r-1}\int^t_{0}erfc(\frac{y}{2\sqrt{s}})ds
+\frac{G_r F}{\sqrt{P_r}(P_r-1)}\int^t_{0}erfc(\frac{y}{2\sqrt{s}})(t-s)ds
\\ \nonumber
&+\frac{G_m\sqrt{S_c}}{S_c-1}\int^t_{0}erfc(\frac{y}{2\sqrt{s}})(t-s)ds
-\frac{G_r}{P_r-1}\int^t_{0}erfc(\frac{\sqrt{P_r}y}{2\sqrt{s}})ds-\frac{G_m}{S_c-1}\int^t_{0}erfc(\frac{\sqrt{S_c}y}{2\sqrt{s}})(t-s)ds
\end{eqnarray}
\subsection{Velocity part corresponding to shear stress for $\bf{\alpha\rightarrow 1}$}
Assuming $\alpha\rightarrow 1$ in shear stress part of Eq. (27), we calculate $u_t(y,t)$
\begin{eqnarray}
u_t(y,t)=L^{-1}\{\bar{u}_t{}(y,q)\}=L^{-1}\{\frac{-F(q)e^{-\sqrt{q+M}y}}{\sqrt{q+M}}\}
\end{eqnarray}
velocity field corresponding to time dependent shear stress and it is
\begin{eqnarray}
u_t(y,t)=-\int^t_0\frac{e^{-\frac{y^2}{4s}-Ms}}{\sqrt{\pi s}}f(t-s)ds
\end{eqnarray}
and
\begin{eqnarray}
\frac{\partial{u_t(y,t)}}{\partial y}=\frac{y}{2\sqrt{\pi}}\int^t_0\frac{e^{-\frac{y^2}{4s}}-Ms}{s^{\frac{3}{2}}}f(t-s)ds
\end{eqnarray}
Eq. (38) can also be written as
\begin{eqnarray}
\frac{\partial{u_t(y,t)}}{\partial y}=\frac{2}{\sqrt{\pi}}\int^\infty_{\frac{y}{2\sqrt{t}}}e^{-s^2-M\frac{y^2}{4s^2}}f(t-\frac{y^2}{4s^2})ds
\end{eqnarray}
Putting $y=0$ in Eq. (39), we obtain
\begin{eqnarray}
\frac{\partial u_t(y,t)}{\partial y}\bigg{|}_{y=0}=f(t)
\end{eqnarray}
satisfying the corresponding boundary condition (14) for velocity $u_t(y,t)$ and as for velocity field $u_{TC}(y,t)$, we observe from Eq. (31)
\begin{eqnarray}
\frac{\partial u_{TC}(y,t)}{\partial y}\bigg{|}_{y=0}=0
\end{eqnarray}
\section{Results and Discussion}
\indent\indent In order to study the influence of fractional parameters $\alpha$, $\beta$, $\gamma$ on motion of MHD fluid over an infinite plate that applies time dependent shear to the fluid, we have drawn several graphs. The effects of physical parameters $G_r$, $G_m$, $S_c$, $P_r$, $M$, $F$ and of fractional parameters $\alpha$, $\beta$, $\gamma$ on free convection of radiative flow will be discussed for the cases of constant shear and oscillating shear applied to fluid.\\
\indent\indent Fig. 1 corresponds to the expression of velocity field for different values of t  and fixed values of $G_r$, $G_m$, $S_c$, $P_r$, $M$, $F$, $\alpha$, $\beta$, $\gamma$ when plate applies constant shear $f(t)=1$ to the fluid. This graph exhibits that velocity is increasing with increasing values of time as well as verifying the boundary condition $u(y,t)\rightarrow 0\,\,\, as\,\,\,y\rightarrow \infty$. Fig. 2 shows velocity profiles for varying parametric values of $G_r$, $G_m$ and $M$ and for fixed values of $S_c$, $P_r$, $F$, $\alpha$, $\beta$, $\gamma$ at $t=0.2$ when plate applies constant shear $f(t)=1$ to the fluid. It is observed that velocity increases with increase in thermal Grashof number $G_r$ and mass Grashof number $G_m$ but it has inverse relation with Hartman number M.\\
\indent\indent In Fig. 3 velocity profiles for different values of $F$ and $M$ are shown at $t=0.1$ and $G_r$, $G_m$, $S_c$, $P_r$, $\alpha$, $\beta$, $\gamma$ are taken to be fixed when plate applies constant shear $f(t)=1$ to the fluid. It is clear from the figure that  velocity increases  with the decrease in $F$ and $M$. Velocity profiles for different values of $S_c$ and $M$ are shown at $t=0.5$ for fixed values of $G_r$, $G_m$, $F$ $P_r$, $\alpha$, $\beta$, $\gamma$ for the case of shear stress $f(t)=1$. This figure depicts the inverse relation of velocity field with $S_c$ and $M$.
\\
\indent\indent In Fig. 5 and Fig. 7 comparison has been made between velocity profiles for different values of $y$ for the cases of constant shear, $f(t)=1$ and oscillating shear, $f(t)=sin(\omega t)$, where $\omega$ is frequency of oscillation. The values of parameters $G_r$, $G_m$, $S_c$, $P_r$, $M$, $F$, $\alpha$, $\beta$, $\gamma$ and  $\omega$ are taken to be fixed. As expected, velocity is decreasing in both cases for increasing value of $y$ because with increasing $y$, the impact of shear applied to fluid by the plate is decreased, reducing the magnitude of velocity. Furthermore, it is observed from Fig. 5 that velocity profiles show oscillating pattern of velocity of viscous MHD fluid owing to oscillating shear. Fig. 7 represents velocity profiles for different values of $t$ and fixed values of $G_r$, $G_m$, $S_c$, $P_r$, $F$, $\alpha$, $\beta$, $\gamma$, $\omega$ when plate applies oscillating shear to the fluid. It can be observed that velocity is increasing with increasing $t$, coinciding with the case of constant shear $f(t)=1$ applied to the fluid, shown in Fig. 1. Also, in Fig. 7, the boundary condition, $u(y,t)\rightarrow 0\,\,\, as\,\,\,y\rightarrow \infty$ is observed to be verified. Another important aspect can also be observed by comparing Fig. 1 and Fig. 7 that velocity in case of oscillating shear has greater magnitude than in the case of constant shear at a particular time $t$.\\
\indent\indent Fig. 9 shows temperature profiles for different values of $F$ and $P_r$ at $t=0.2$ and fixed value of fractional parameter $\beta$. It can be seen that temperature increases with decreasing values of $F$ and $P_r$. However, a rapid change is observed in temperature profiles for the values of $P_r=0.7$ and $P_r=5$ i.e. temperature decreases sharply for rapid increase in Prandtl number $P_r$. In Fig. 10, we observe the behavior of mass concentration of fluid for different values of $S_c$ at $t=0.2$ and fixed value of fractional parameter $\gamma$. It clearly shows that increasing Schmidt number, $S_c$ has negative impact on concentration of MHD fluid and vice versa.\\
\indent\indent Lastly, Fig. 8, Fig. 10 and Fig. 11 reflects on the influence of fractional parameters  $\alpha$, $\beta$, $\gamma$ on velocity, temperature and mass concentration of fluid, respectively, for the case of constant shear $f(t)=1$ applied to the fluid by infinite plate. These figures show that for decreasing values of parameters $\alpha$, $\beta$ and $\gamma$, velocity, temperature and mass concentration of fluid increase, respectively. Also, we retrieve profiles of velocity, temperature and concentration for governing equations with ordinary differential operators by taking $\alpha,\,\,\beta,\,\,\gamma\rightarrow 1$.\\
\indent\indent To establish the validity of analytical solutions, the numerical results for concentration have been prepared. A comparison of values of concentration obtained by using Stehfest's numerical algorithm [58] for calculating inverse Laplace transform of Eq. (19) has been made with the values of concentration calculated from Eq. (20) for $n=55$ terms. These results are shown in Table 1. Stehfest's algorithm is defined by followinng relation
\begin{equation}
C(y,t)=L^{-1}\{\bar{C}(y,q)\}\approx\frac{\ln2}{t}\sum_{k=1}^{2r}b_k\bar{C}\bigg(y,k\frac{\ln2}{t}\bigg),
\end{equation}
where $r$ is a positive integer,
\begin{equation}
b_k=(-1)^{k+r}\sum_{s=\bigg[\frac{k+1}{2}\bigg]}^{min(k,r)}\frac{s^r(2s)!}{(r-s)!s!(s-1)!(k-s)!(2s-k)!}
\end{equation}
and $[p]$ denotes the integer part of the real number p. Table 1 shows the accuracy of analytical results upto the order of $10^{-6}$, thus validating our solutions of concentration. Similar approach could be adopted for validation of velocity and temperature results.\\\\
\textbf{Table 1.}
\textbf{Values of concentration C(y, t) resulting from the analytic solution Eq. (20) and the numerical algorithm applied to Eq. (19) at t = 5, Sc = 1 and
$\gamma= 0.58$}
\begin{center}
\begin{tabular}{|c|c|c|c|}
%{||C(y,t)-Eq.(20)|C(y,t)-Eq.(19)|Absolute Error|}
\hline
 $\textbf{y}$ & $\textbf{C(y,t)-Eq.(20)}$ & $\textbf{C(y,t)-Eq.(18)}$ & $\textbf{Absolute Error}$\\
 \hline
 0 & 5 & 5.00001 & $6.031\times10^{-6}$\\
 \hline
 0.1 & 4.66656 & 4.66658 & $1.694\times10^{-5}$\\
 \hline
  0.2 & 4.35403 & 4.35404 & $1.338\times10^{-5}$ \\
\hline
  0.3 & 4.06119 & 4.0612 & $3.346\times10^{-6}$ \\
\hline
  0.4 & 3.7869 & 3.78691 & $6.425\times10^{-6}$ \\
\hline
  0.5 & 3.53007 & 3.53008 & $5.33\times10^{-6}$ \\
\hline
  0.6 & 3.28967 & 3.28968 & $4.329\times10^{-6}$ \\
\hline
  0.7 & 3.06472 & 3.06473 & $6.388\times10^{-7}$ \\
\hline
  0.8 & 2.8543 & 2.8543 & $5.092\times10^{-6}$ \\
\hline
  0.9 & 2.65753 & 2.65754 & $6.509\times10^{-6}$ \\
\hline
  1 & 2.47359 & 2.47359 & $2.626\times10^{-6}$ \\
\hline
\end{tabular}
\end{center}

\section{Conclusion}
\indent\indent Theoretical study of an MHD viscous fluid flow over an infinite plate that applies an arbitrary shear to the fluid has been performed and an analysis of flow-enhancing and flow-hindering factors has been prepared. Caputo fractional differential operator owing the efficiency of fractional constitutive equations to assess information about molecular movement of particles as well as for the purpose of generalization has been employed. Exact expressions of velocity field, temperature and mass concentration have been obtained by taking Laplace transform of dimensionless fractional differential equations. Both temperature and mass concentration are represented in series solutions. The velocity of fluid is expressed as a sum of two functions i.e. one corresponds to the shear applied by the plate to fluid and the second part corresponds to thermal radiation and mass concentration. We observe that initial and boundary conditions for temperature and mass concentration are verified by the obtained solutions directly. However, for the case of velocity field we verified the shear stress boundary condition for the shear $f(t)=1$. It could also be observed that the part of velocity corresponding to shear stress vanishes if plate is applying no stress to the fluid i.e. $f(t)=0$\\
\indent\indent Some significant limiting cases of fluid parameters and fractional parameters have also been discussed. Particularly, $\beta\rightarrow 1$ and thermal radiation parameter, $F=0$ lead to known expression of temperature in the form of complementary error function. Some interesting facts are established by considering two cases of shear stress. The first case corresponds to the motion of fluid over an infinite plate that applies constant shear $f(t)=1$ to the fluid and in the second case, plate applies oscillating shear to the fluid. Finally, some physical aspects of fluid motion are brought to light through graphs. Graphical findings are summed up as follows:\\
\textbf{1.}  Both temperature and mass concentration of fluid are independent of shear being applied to fluid by the plate. Temperature has inverse relation with Prandtl number, $P_r$ and thermal radiation parameter, $F$. As increase in temperature will cause increase in thermal radiation emission that will ultimately decrease the temperature.  Also, concentration has inverse relation with Schmidth number, Sc.\\
\textbf{2.} Velocity increases with the increase in the parametric values of thermal Grashof number, $G_r$ and mass Grashof number, $G_m$ and decreases with the increase of parameters $S_c$, $F$ and $M$.\\
\textbf{3.} Comparison of velocity profiles for both cases of shear stress verifies the fact that velocity decreases for increasing values of $y$. This decrease in velocity is due to less impact of shear induced on fluid being away from the plate at height. A significant difference between graphs of velocity for two cases is observed as an oscillating pattern of velocity profiles for the case of oscillating shear distinguishes it from the other case of constant shear stress.\\
\textbf{4.} The influence of fractional parameters on fluid motion is also depicted through graphs. It is observed that velocity, temperature and concentration decrease with increasing values of parameters $\alpha$, $\beta$ and $\gamma$ for the case of constant shear applied to fluid by the plate.\\
\textbf{5.} We have obtained numerical solutions for concentration of fluid and have compared these solutions with analytical solutions. The accuracy and validity of analytical solutions have been established by calculating absolute error being of order $10^{-6}$. Such validity could be checked for temperature and velocity.\\
\indent\indent The graphical results are in direct correspondence with physical understanding of the model as the added force of shear applied by the plate on fluid increases the mass and thermal diffusivity near the boundary and thus increasing the thermal Grashof number, $G_r$ and mass Grashof number, $G_m$. The increased turbulence near boundary causes increase in velocity of fluid. However, magnetic forces and collective movement of fluid particles impede smooth flow of fluid over plate, causing decrease in velocity of fluid. This fact is illustrated in graphs that increase in Schmidt number, $S_c$ and Hartmann number, $M$ cause decrease in velocity.
\\
{\textbf{Appendix}}
\begin{equation*}
L^{-1}\{q^\alpha e^{-uq^\alpha}\}=\int_0^\infty J_0(2\sqrt{xt})\frac{1}{\alpha}\sum_{n=0}^\infty\frac{(-u)^n x^{\alpha(n+1)}}{(n+1)!\Gamma{[\alpha(n+1)]}}, \hskip2.3cm (A_1)
\end{equation*}
\begin{equation*}
L^{-1}\{\frac{q^b}{(q^a-d)^c}\}=G_{a,b,c}(d,t);\,\,\,\,Re(ac-b)>0,\,\,Re(q)>0,\,\,|\frac{p}{q^a}|<1, \hskip0.3cm (A_2)
\end{equation*}
\begin{equation*}
\frac{1}{P_r q^\beta-q^\alpha+(F-M)}=\frac{1}{P_r}\sum_{p=0}^\infty\frac{(\frac{1}{P_r})^p q^{p\alpha}}{(q^\alpha-\frac{M-F}{P_r})^{p+1}}, \hskip3.9cm (A_3)
\end{equation*}
\begin{equation*}
\frac{1}{S_cq^\gamma-q^\alpha-M}=\frac{1}{S_c}\sum_{m=0}^\infty\frac{(\frac{1}{S_c})^m q^{m\alpha}}{(q^\gamma-\frac{M}{S_c})^{m+1}}, \hskip5.4cm (A_4)
\end{equation*}
\begin{equation*}
L^{-1}\bigg\{\frac{e^{-\sqrt{q+M}y}}{\sqrt{q+M}}\bigg\}=e^{-Mt}\bigg(\frac{e^{-\frac{y^2}{4t}}}{\sqrt{\pi t}}\bigg), \hskip6.5cm (A_5)
\end{equation*}
\begin{equation*}
L^{-1}\bigg\{\frac{e^{-\sqrt{P_rq}y}}{q}\bigg\}=erfc\bigg(\frac{\sqrt{p_r}y}{2\sqrt{t}}\bigg), \hskip6.6cm (A_6)
\end{equation*}

{\textbf{References}
\begin{description}

\item{[1]} M. A. Hossain, H. S. Takhar, Radiation effect on mixed convection along a vertical plate with uniform surface temperature, Heat Mass Transf., 31(4) (1996), 243-248
\item{[2]} K. Vajravelu, A. Hadjiricaloou , Convective heat transfer in an electrically conducting fluid at a stretching surface with uniform free stream, Int. J. Eng. Sci., 35 (1997), 1237-1244 .
\item{[3]} M. A. El-Hakiem, Radiation effects on hydromagnetic free convective and mass transfer flow of a gas past a circular cylinder with uniform heat and mass flux, Int. J. Numer. Method. H., 19 (3/4) (2009), 445-458.
\item{[4]} M. A. Samad, M. Mohebujjaman, MHD heat and mass transfer free convection flow along a vertical stretching sheet in presence of magnetic field with heat Generation, Res. J. Appl. Sci. Eng. Tech., 1(3) (2009), 98-106.
\item{[5]} S. T. Khaleque, M. A. Samad, Effects of radiation, heat generation and viscous dissipation on MHD free convection flow along a stretching sheet, Res. J. Appl. Sci. Eng. Tech., 2(4) (2010), 368-377.
\item{[6]} I. Zahan, M. A. Samad, Radiative heat and mass transfer of an MHD free convection flow along a stretching sheet with chemical reaction, heat generation and viscous dissipation, Dhaka Univ. J. Sci., 61(1) (2013), 27-34.
\item{[7]} M. K. Nayak, G. C. Dash, L. P. Singh, Unsteady radiative MHD free convective flow and mass transfer of a viscoelastic fluid past an inclined porous plate, Arab. J. Sci. Eng., 40(11) (2015), 3029-3039.
\item{[8]} S. A. Sapareto, W. C. Dewey, Thermal dose determination in cancer therapy, Int. J. Radiat. Oncol., 10(6) (1984), 787-800.
\item{[9]} A. K. Datta, Porous media approaches to studying simultaneous heat and mass transfer in food processes. I: Problem formulations, J. Food Eng., 80(1) (2007), 80-95.
\item{[10]} Srinivasacharya, Darbhashayanam, U. Mendu, Thermal radiation and chemical reaction effects on magnetohydrodynamic free convection heat and mass transfer in a micropolar fluid, Turk. J. Eng. Environ. Sci., 38 (2014), 184-196.
\item{[11]} C. Zhang, L. Zheng, X. Zhang, G. Chen, , MHD flow and radiation heat transfer of nanofluids in porous media with variable surface heat flux and chemical reaction, Appl. Math. Model., 39(1) (2015), 165-181.
\item{[12]} R. Kandasamy,  I. Hashim,  Muhaimin, Seripah, Nonlinear MHD mixed convection flow and heat and mass transfer of first order chemical reaction over a wedge with variable viscosity in the presence of suction or injection. Theoret. Appl. Mech., 34(2) (2007), 111-134.
\item{[13]} Z. Uddi, M. Kumarb, MHD heat and mass transfer free convection flow near the lower stagnation point of an isothermal cylinder imbedded in porous domain with the presence of radiation, Jord. J. Mech. Ind. Eng.. 5(2) (2011), 133-138.
\item{[14]} C. Huang , Y. Zhang, Calculation of high-temperature insulation parameters and heat transfer behaviors of multilayer insulation by inverse problems method, Chinese J. Aeron., 27(4) (2014), 791-796.
\item{[15]} M. M. Rashid, B. Rostami, N. Freidoonimehr, S. Abbasbandy, Free convective heat and mass transfer for MHD fluid flow over a permeable vertical stretching sheet in the presence of the radiation and buoyancy effects, Ain Shams Eng. J., 5(3) (2014), 901-912.
\item{[16]} G. S. Seth, S. M. Hussain, S. Sarkar, , Hydromagnetic natural convection flow with heat and mass transfer of a chemically reacting and heat absorbing fluid past an accelerated moving vertical plate with ramped temperature and ramped surface concentration through a porous medium, J. Egyp. Math. Soc., 23(1) (2015), 197-207.
\item{[17]}  L. Wu, Mass transfer induced slip effect on viscous gas flows above a shrinking/stretching sheet. Int. J. Heat Mass Tran., 93 (2015), 17-22.
\item{[18]} R. S. Sapieszko,E. Matijevi, Preparation of well-defined colloidal particles by thermal decomposition of metal chelates. I. Iron oxides, J. Collo. Interf. Sci., 74(2) (1980), 405-422.
\item{[19]} S. Taniguchi, A. Kikuchi , Flow control of liquid iron by magnetic shield in a high-frequency induction furnace, Tetsu-to-Hagane, 78(5) (1992), 753-760
\item{[20]} C. W. Forsberg, Hydrogen, nuclear energy and the advanced high-temperature reactor, Int. J. Hydro. Energy, 28(10) (2003), 1073-1081.
\item{[21]} B. Yildiz, M. S. Kazimi, Efficiency of hydrogen production systems using alternative nuclear energy technologies, Int. J. Hydro. Energy. 31(1) (2006), 77-92.
\item{[22]} V. M. Soundalgekar, Free convection effects on the flow past a vertical oscillating plate, Astrop. Space Sci., 66 (1979), 165-172.
\item{[23]} V. M. Soundalgekar and S. P. Akolkar, Effects of free convection currents and mass transfer on flow past a vertical oscillating plate, Astrop. Space Sci., 89 (1983) 241-254.
\item{[24]} V. M. Soundalgekar, R. M. Lahurikar, S. G. Pohanerkar and N. S. Birajdar, Effects of mass transfer on the flow past an oscillating infinite vertical plate with constant heat flux, Thermoph. Aeromech., 1 (1994), 119-124.
\item{[25]} W. G. England, A. F. Emery, Thermal radiation Effects on the laminar free convection boundary layer of an absorbing gas, J. Heat Trans., 91 (1969), 37-44.
\item{[26]} P. S. Gupta, A. S. Gupta, Radiation Effect on hydromagnetic convection in a vertical channel, Int. J. Heat Mass Trans., 17 (1974), 1437-1442.
\item{[27]} M. A. Hossain, H. S. Takhar, Radiation Effect on mixed convection along a vertical plate with uniform surface temperature, Heat Mass Transf., 31 (1968), 243-248.
\item{[28]} M. K. Mazumdar, R. K. Deka, MHD flow past an impulsively started infinite vertical plate in presence of thermal radiation, Rom. J. Phy., 52(5-6) (2007), 529-535.
\item{[29]} B. Gebhart, L. Pera, The nature of vertical natural convection flows resulting from the combined buoyancy effects of thermal and mass diffusion, Int. J. Heat Mass Trans., 14 (1971), 2025-2050.
\item{[30]} R. K. Deka, B. C. Neog, Combined effects of thermal radiation and chemical reaction on free convection flow past a vertical plate in porous medium, Adv. Appl. Fluid Mech., 6 (2009), 181-195.
\item{[31]} N. D. Waters and M. J. King, Unsteady flow of an elastico-viscous liquid, Rheol. Acta., 9 (1970), 345-355.
\item{[32]} R. Bandelli, K. R. Rajagopal, G. P. Galdi, On some unsteady motions of fluids of second grade, Arch. Mech., 47, 4 (1995), 661–676,
\item{[33]} Y. Yao, Y. Liu, Some unsteady flows of a second grade fluid over a plane wall, Nonlinear Anal. RWA, 11 (2010), 4442–4450.
\item{[34]} D. Vieru, C. Fetecau, A. Sohail, Flow due to a plate that applies an accelerated shear to a second grade fluid between two parallel walls perpendicular to the plate, Z. Angew. Math. Phys., 62 (2011), 161-172.
\item{[35]} C. Fetecau, D. Vieru, Corina Fetecau, Effect of side walls on the motion of a viscous fluid induced by an infinite plate that applies an oscillating shear stress to the fluid, Cent. Eur. J. Phys., 9(3) (2011), 816-824.
\item{[36]}  M. Seredy´nska, A. Hanyga, Nonlinear differential equations with fractional damping with applications to the 1dof and 2dof pendulum, Acta Mech., 176 (2005), 169-183.
\item{[37]}  D. Baleanu, S. I. Muslih, K. Tas, Fractional Hamiltonian analysis of higher order derivatives systems, J. Math. Phys. 47, 10 (2006) art. no. 103503, 8 .
\item{[38]} A. Carpinteri, F. Mainardi, Fractals and fractional calculus in continuum mechanics, Springer, New York, 1997.
\item{[39]} O. P. Agrawal, Solution for a fractional diffusion-wave equation defined in a bounded domain, Nonlinear Dynam., 29 (2002), 145-155.
\item{[40]} A. Germant, On fractional differentials, philosophical Magazine, 25 (1938), 540-549.
\item{[41]} G. L. Slonimsky, Laws of mechanical relaxation processes in polymer, J. Polym. Pol. Sym., 16 (1967), 1667-1672.
\item{[42]} R. L. Bagley, P. J. Torvik, Fractional Calculus-differential approach to the analysis of viscoelasticity damped structures, AIAA J., 21 (1983), 742-748.
\item{[43]} R. L. Bagley, P. J. Torvik, A theoretical basis for the applications of fractional calculus of viscoelasticity, J. Rheol., 27 (1983), 201-210.
\item{[44]} R. C. Koeller, Applications of fractional calculus to the theory of viscoelasticity, J. Appl. Mech., 51 (1984), 299-307.
\item{[45]} S. Wang, M. Xu, Axial Coutte flow of two kinds of fractional viscoelastic fluids in an annulus, Nonlinear Anal. Real World Appl., 10 (2009), 1087-1096.
\item{[46]} D. Tripathi, S. K. Pandey, S. Das, Peristaltic flow of viscoelastic fluid with fractional Maxwell model through a channel, Appl. Math. Comput., 215 (2010), 3645–3654.
\item{[47]} T. Hayat, S. Zaib, C. Fetecau, Corina Fetecau, Flows in a fractional generalized Burgers' fluid, J. Porus Media, 13 (2010), 725-739.
\item{[48]} T. Hyat, S. Najam, M. Sjid, M. Ayub, S. Mesloub, On exact solutions for oscillatory flows in a generalized Burgers' fluid with slip condition, Z. Naturforsch, 65a (2010) 381-391.
\item{[49]} Y. Liu, L. Zheng, Unsteady MHD Couette flow of a generalized Oldroyd-B fluid with fractional derivative, Comput. Math. Appl., 61 (2011), 443-450.
\item{[50]} C. Fetecau, Corina Fetecau, M. Jamil, A. Mahmood, Flow of fractional Maxwell fluid between coaxial cylinders, Arch. App. Mech., 81 (2011), 1153-1163.
\item{[51]} D. Tripathi, P. K. Gupta, S. Das, Influence of slip condition on peristaltic tansport of aviscoelastic fluid with model, Therm. Sci., 15 (2011), 501-515.
\item{[52]} M. Jamil, A. Rauf, A. A. Zafar, N. A. Khan, New exact analytical solutions for Stokes' first problem of Maxwell fluid with fractional derivative approach, Comput. Math. Appl., 62 (2011), 1013-1023.
\item{[53]} M. Jamil, N. A. Khan, N. Shahid, Fractional MHD Oldroyd-B fluid over an oscillating plate, Therm. Sci., Vol. 17, No. 4 (2013), 997-1011.
\item{[54]} N. Shahid, A study of heat and mass transfer in a fractional MHD flow over an infinite oscillating plate, SpringerPlus, 04, 01,doi: 10.1186/s40064-015-1426-4 (2015).
\item{[55]} I. Podlubny, Fractional differential equations, Academic press, San Diego, (1999).
\item{[56]} F. Mainardi, Fractional calculus and waves in viscoelasticity: An introduction to mathematical models, Imp. College Press, London (2010).
\item{[57]} A. M. Mathai, R. K. Saxena, H. J. Haubold, The H-functions: Theory and Applications, Springer, New York (2010).
\item{[58]} H. Stehfest, Algorithm 368: numerical inversion of Laplace transform, Communication of the ACM, 13 (1970) 47-49.
\end{description}
\begin{figure}[]
\includegraphics[width=14cm, height=20cm]{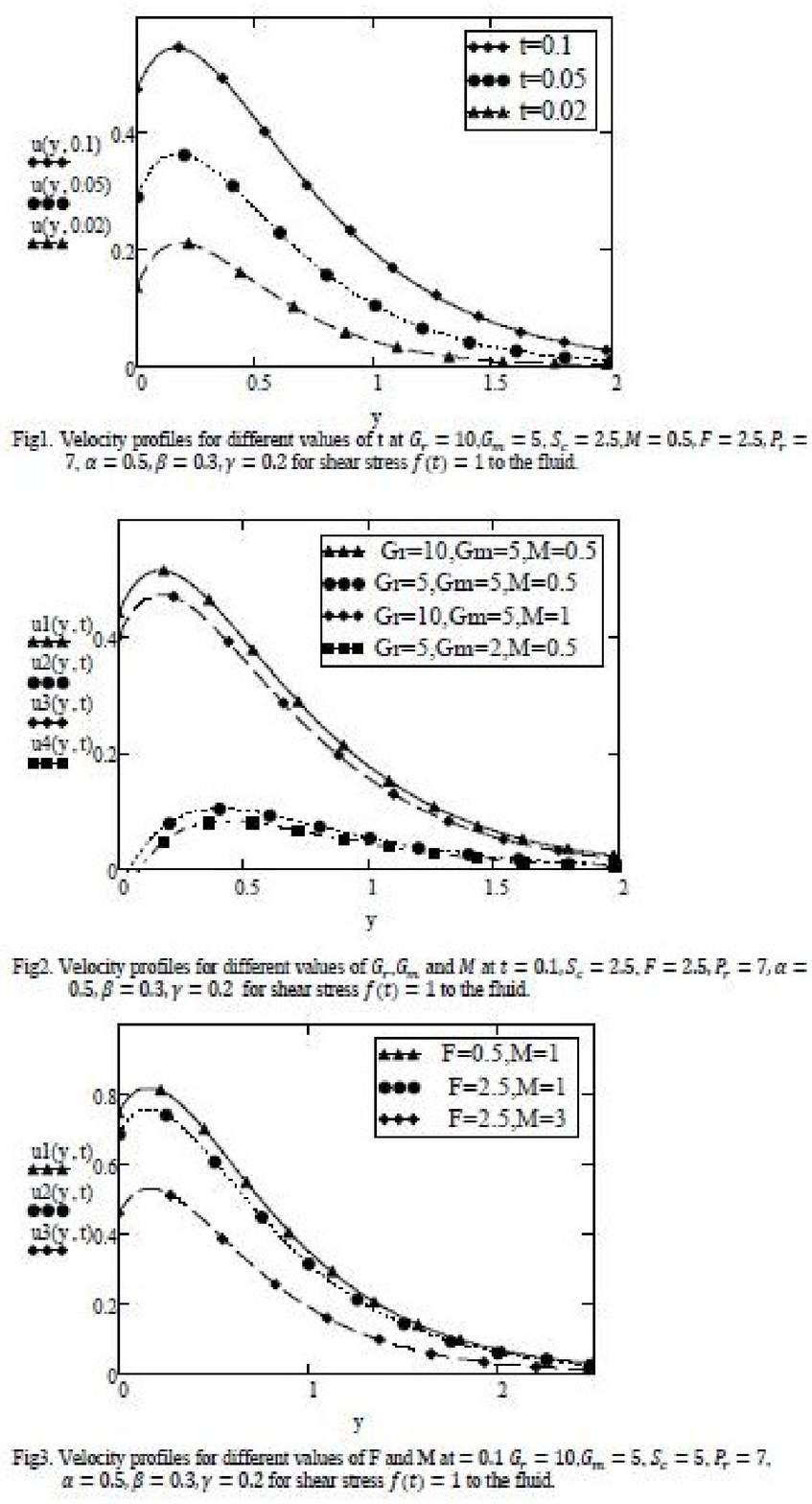}
\end{figure}
\begin{figure}[]
\includegraphics[width=14cm, height=20cm]{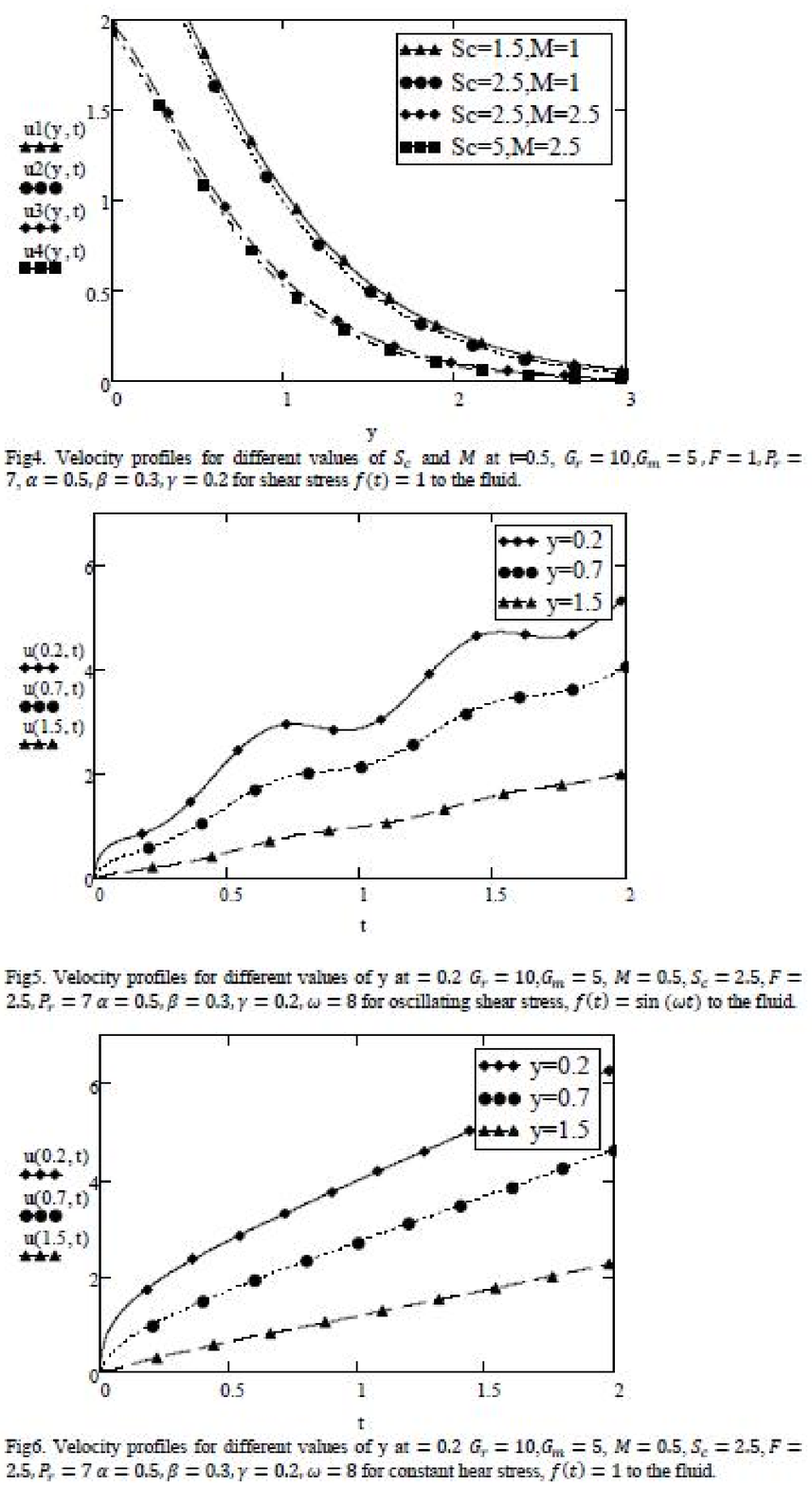}
\end{figure}
 \begin{figure}[]
 \includegraphics[width=14cm, height=20cm]{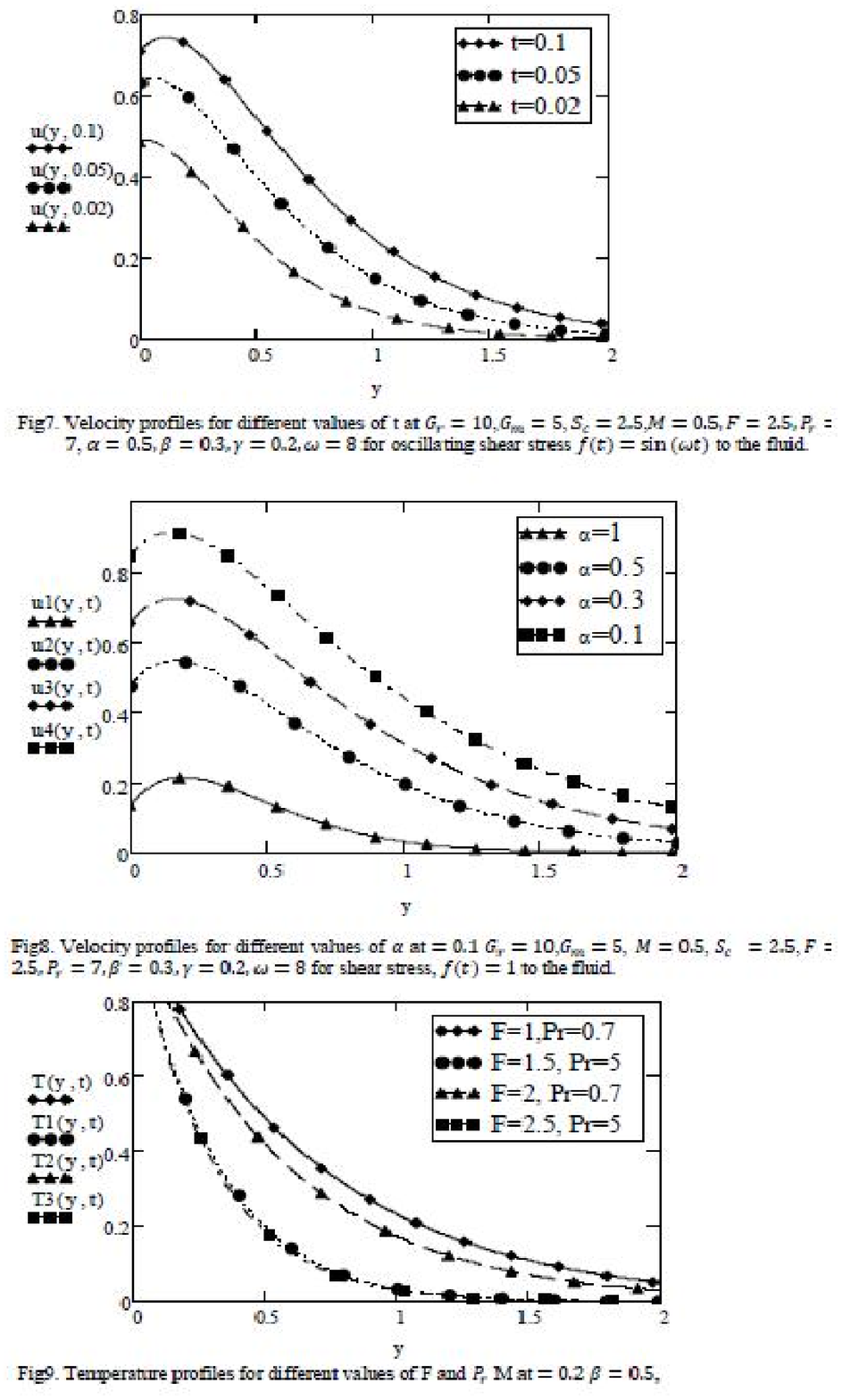}
 \end{figure}
 \begin{figure}[]
\includegraphics[width=14cm, height=20cm]{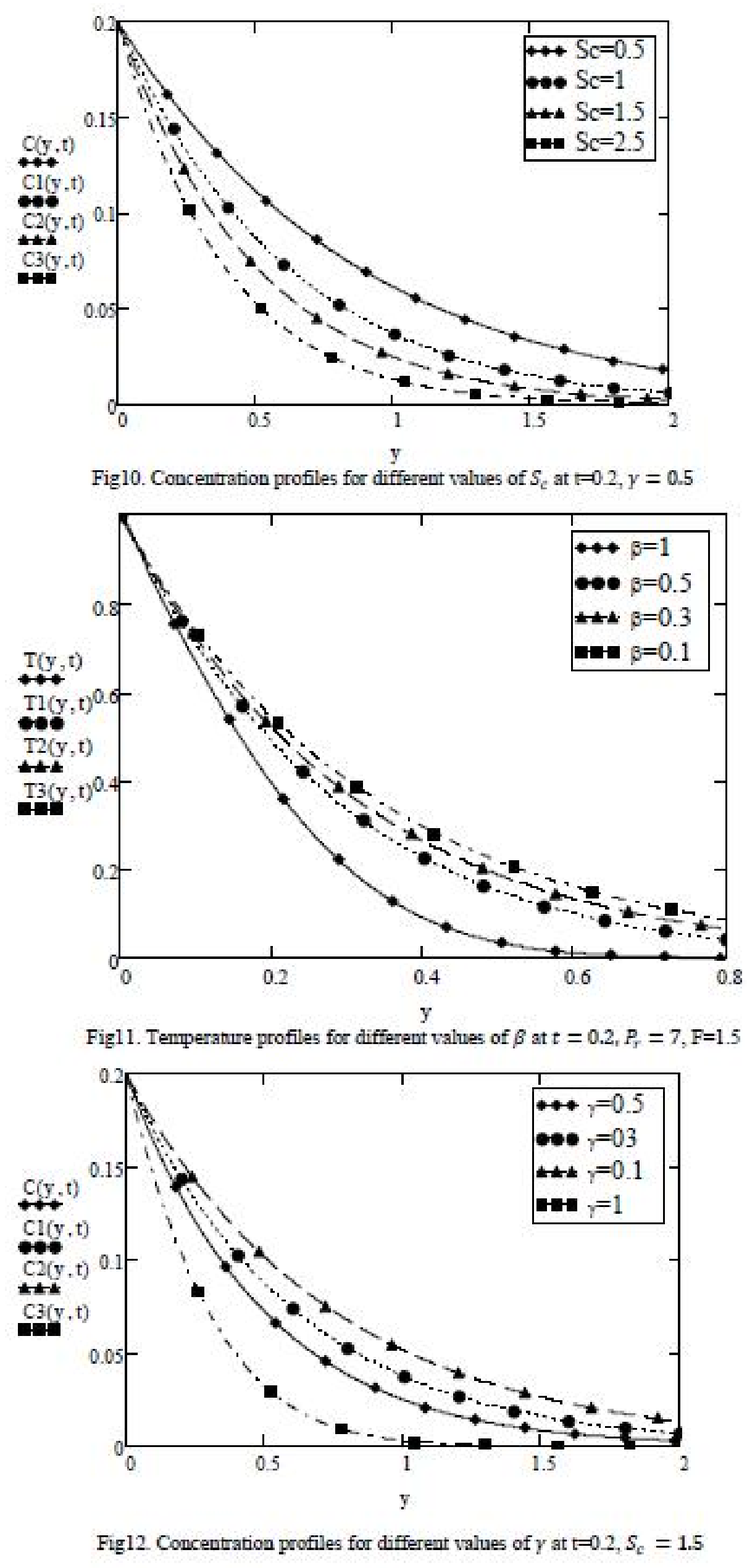}
 \end{figure}
 \end{document}